\documentclass[letterpaper]{article} 
\usepackage{aaai23}  
\usepackage{times}  
\usepackage{helvet}  
\usepackage{courier}  
\usepackage[hyphens]{url}  
\usepackage{graphicx} 
\usepackage{natbib}  
\usepackage{caption} 
\usepackage{algorithm}
\usepackage{algorithmic}

\usepackage{indentfirst}
\usepackage{enumerate}
\usepackage{lscape, amsbsy, amstext}
\newcommand{\T}{\top}

\usepackage{amstext}
\usepackage{amssymb}
\usepackage{amsmath}
\usepackage{caption}
\usepackage{subcaption}
\usepackage{fancyhdr}

\def\be{\begin{equation}} 
  \def\ee{\end{equation}} 
  \def\beqn{\begin{eqnarray}} 
  \def\eeqn{\end{eqnarray}} 
  \def\beq{\begin{eqnarray*}} 
  \def\eeq{\end{eqnarray*}}

\newtheorem{theorem}{Theorem}

\newtheorem{proposition}{Proposition}
\newtheorem{Example}{Example}

\usepackage{mathrsfs}
\usepackage{lineno}
\usepackage{amsfonts}

\usepackage{newfloat}
\usepackage{listings}
\DeclareCaptionStyle{ruled}{labelfont=normalfont,labelsep=colon,strut=off} 
\floatstyle{ruled}
\newfloat{listing}{tb}{lst}{}
\floatname{listing}{Listing}
\title{Spearman Rank Correlation Screening for Ultrahigh-dimensional Censored Data}
\author{
    Hongni Wang$^1$,
    Jingxin Yan$^2$,
    Xiaodong Yan$^*$$^3$
}
\affiliations{
    \textsuperscript{\rm 1}School of Statistics and Mathematics, Shandong University of Finance and Economics\\
    \textsuperscript{\rm 2}Academy of Mathematics and Systems Science,
Chinese Academy of Sciences\\
    \textsuperscript{\rm 3}Zhongtai Securities Institute for Financial Studies, Shandong University\\

    yanxiaodong@sdu.edu.cn
}

\usepackage{bibentry}

\begin{document}

\maketitle

\begin{abstract}
Herein, we propose a Spearman rank correlation based screening procedure for ultrahigh-dimensional data with censored response case.
The proposed method is model-free without specifying any regression forms of predictors or response variable and is robust under the unknown monotone transformations of these response variable and predictors. The sure-screening and rank-consistency properties are established under some mild regularity conditions. Simulation studies demonstrate that the new screening method  performs well in the presence of a  heavy-tailed distribution, strongly dependent predictors or outliers and that offers
 superior performance over the existing nonparametric screening procedures. In particular, the new screening method still works well when a response variable is observed under a high censoring rate. An illustrative example is provided.
\end{abstract}


\section{Introduction}

Ultrahigh-dimensional covariates  are often encountered in many fields of study, e.g.,  mechanical systems, genetic engineering \cite{zhou2021category}, and biomedical engineering. Under the ``larger $p$ smaller $n$'' data framework, numerous penalized variable selection approaches have been developed for high-dimensional Cox model \cite{zhang2007adaptive, zou2008note}, additive hazard model \cite{chen2018reweighted,leng2007path,martinussen2009covariate,lin2013high},  linear regression model \cite{huang2008asymptotic,wang2008doubly} and unconditional moment model \cite{tang2018exponentially} . However, the aforementioned variable selection methods may not perform well because of the simultaneous challenges of computational expendiency, statistical accuracy,
sophisticated algorithm techniques, and strong model assumptions \cite{fan2009ultrahigh, tang2018exponentially}. Therefore, it is necessary and urgent to develop new approaches to deal with ultrahigh-dimensional censored data.

Recently, emerging feature screening approaches have been developed for a complete response with ultrahigh-dimensional covariates. The existing model-based feature-screening methods include sure independence screening (SIS) for linear regression \cite{fan2008sure}, a maximum-marginal-likelihood (MML) approach for generalized linear models \cite{fan2010sure}, nonparametric feature screening method for missing response \cite{li2020nonparametric} nonparametric independence screening (NIS) for additive models \cite{fan2011nonparametric},  partially linear models with missing responses \cite{tang2019feature}, and marginal empirical-likelihood screening (ELS) for linear regression \cite{chang2013marginal}.  To overcome the limitations of imposing working models,  model-free feature-screening methods have been developed. These include sure independent ranking and screening (SIRS) \cite{zhu2011model},  rank-correlation screening (RCS) \cite{li2012robust}, distance correlation screening (DCS) \cite{li2012feature, chen2019note}, category-adaptive variable screening \cite{xie2020category}, quantile-adaptive screening (QAS) \cite{he2013quantile}, fused Kolmogorov filter screening (FKFS) \cite{mai2015fused}, and conditional quantile screening (CQS) \cite{wu2015conditional}, and the fused mean-variance (FMV) filter \cite{yan2018fused}.

However, the aforementioned screening procedures proposed for a complete response failure in performing well for the censored model because it cannot be estimated reliably. In particular, the performance will be dramatically decreased under a high censoring rate or in the existence of  outliers in the predictors.
 Furthermore, several emerging feature-screening procedures have been proposed for censored responses. The existing model-based screening methods mainly focus on the Cox model. These include the lasso-penalization
approach for prescreening \cite{tibshirani1997lasso}, the standardized-marginal-maximum partial-likelihood estimators \cite{zhao2012principled}, and the marginal-sure-independence-screening procedure \cite{fan2010sure}. However, model-free screening methods with a censored response, such as the censored quantile-adaptive screening procedure \cite{he2013quantile},  the censored-rank-correlation screening (RCS$_{\rm cen}$) procedure with  inverse
probability-of-censoring weighted as Kendall's $\tau$ \cite{song2014censored}, the conditional-quantile screening (CQS$_{\rm cen}$) procedure for a covariate-independent censoring method \cite{wu2015conditional},
the adjusted-distance-orrelation  screening procedure (DCS$_{\rm cen}$) \cite{chen2018robust},
and the censored  sure  independent ranking and screening methods (SIRS$_{{\rm cen}}$) \cite{zhou2017model}, may be more robust under model misspecification. Recent research on the screening issue of ultrahigh-dimensional censored data has also included the works of \cite{liu2018quantile,lin2018model,liu2018new,zhang2017correlation,zhang2018censored}.

The main contributions of this article include the followings:
\begin{itemize}
\item This is the attempt using imputation technic to adjust the feature screening method in high-dimensional censored data, called as Spearman rank correlation screening (SRCS$_{\rm cen}$).
\item It is model-free due to it is invariant under monotonic transformations of the response and robust in the presence of monotonic transformations of predictor variables.
    \item The feature screening performance still behaves under a high censoring rate.
\end{itemize}

\section{Screening Procedures}

Let $Y$ be a continuous response with a support $\mathbb{R}^1_y$, and $\boldsymbol{X}$ be a vector of continuous covariates with a support $\mathbb{R}^p_{\boldsymbol{x}}$. Define $F_k(x)= {\rm pr}(X_k\le x)$, $F_{Y|X_k}(y\mid x)= {\rm pr}(Y\le y\mid X_k=x)$, $F_k(x,y)= {\rm pr}(X_k\le x, Y\le y)$
($k=1, \ldots, p$), and $F(y)= {\rm pr}(Y\leq y)$. To investigate the relationship between $X_k$ and $Y$, Fan and Lv \cite{fan2008sure} utilized absolute marginal Pearson correlation $E(X_kY)-E(X_k)E(Y)$ to rank the linear correlation between $X_k$ and $Y$. Zhu et al. \cite{zhu2011model} proposed a marginal screening utility based on $E\{X_kF_{Y|X_k}(Y\mid X_k)\}$, while  Li et al. \cite{li2012robust} developed  the robust rank correlation screening method based on $E\{F_k(X_k, Y)\}-E\{F_k(X_k)\}E\{F(Y)\}$. Motivated by these marginal screening utilities, we consider investigating the correlation between the distributions $F_k(x)$   and $F(y)$, because such correlations include the linear and nonlinear relationship between $X_k$ and $Y$. Therefore, we propose the following index for the $k$th covariate,
\begin{align}
\label{SRCS}
\nonumber
\omega_k = E\{F_k(X_k)F(Y)\}-E\{F_k(X_k)\}E\{F(Y)\} = \\ \iint\{F_k(x)F(y)\}dF_k(x,y)-\frac{1}{4},
\end{align}
which can be used to measure the dependence between $X_k$ and $Y$. Then, $\omega_k$ serves as the population quantity of our proposed
marginal-utility measure for ranking the potential correlations between predictors and responses. $\omega_k$ has the remarkable property of being 0 if  $X_k$ and $Y$ are statistically independent. This motivates us to utilize it for feature-screening to characterize both the linear and nonlinear relationships between  responses and  covariates.

Suppose that one observes the right-censored  data  $(\boldsymbol{X}, Y^*, \delta)$=$\{\boldsymbol{X},Y\land C$, $I(Y\leq C)\}$, where $Y$ is  a censored response of interest, $C$ represents a censoring variable, and
${\boldsymbol{ X}}=(X_{1}, \ldots, X_{p})^T$. Let $\{\boldsymbol{X}_i, Y^*_i,\delta_i: i=1, \ldots, n\}$  comprise independent copies of $(\boldsymbol{X}, Y^*, \delta)$.
 If we know the distributions $F_k(x)$ and $F(y)$, the moment estimator of $E\{F_k(x)F(y)\}$ in (\ref{SRCS}) is   $1/n\sum_{i=1}^n\{F_k(X_{ki})F(Y_i)\}$. Since $Y_i$ cannot be completely observed due to censoring, we replace $F(Y_i)$ with its conditional expectation given the observed response $Y_i^*$, censoring indicator $\delta_i$ and predictor $X_{ki}$. 
 For this, let
 $\mathcal{F}(Y^*_i,\delta_i,X_{ki})=E\{F(Y_i)\mid Y_i^*,\delta_i, X_{ki}\}$. Note that
\begin{small}
\begin{eqnarray*}\label{Imput}
\mathcal{F}(Y^*_i,\delta_i,X_{ki})
  \!&\!=\!&\delta_iF(Y^*_i)\!+\!(1\!-\! \delta_i)E\{F(Y_i)\mid Y_i \!>\! Y_i^*,\! Y^*_i,\! X_{ki}\}\\
 \! &\!=\!&\delta_iF(Y^*_i)\!+\!(1\!-\! \delta_i)\frac{E\{F(Y_i)I(Y_i>Y_i^*)\mid X_{ki}\}}{1\!-\!F(Y_i^*\mid X_{ki})}\\
 \!& \!=\! &\delta_iF(Y^*_i)\!+\!(1\!-\! \delta_i)\frac{\int_{Y^*_i}^\infty F(y)dF(y\mid X_{k_i})}{1-F(Y^*_i\mid X_{k_i})},
\end{eqnarray*}
\end{small}
where the equality is derived from the formula of conditional expectation given a event, $F(y\mid x)$ is the $Y$ distribution given the fixed value $x$. Then we conclude that
\begin{proposition}\label{prop1}
 The unbiased moment estimator of $E\{F_k(x)F(y)\}$ is   $1/n\sum_{i=1}^n\{F_k(X_{ki})\mathcal{F}(Y^*_i,\delta_i,X_{ki})\}$. i.e., $E\{F_k(X_{ki})\mathcal{F}(Y^*_i,\delta_i,X_{ki})\}=E\{F_k(x)F(y)\}$.
\end{proposition}

 However, the distributions of $F_k(x)$, $F(y)$ and $F(y\mid x)$  are usually unknown in practice. $F_k(x)$ can be estimated by its empirical distribution $\widehat{F}_k(x)=1/n \sum_{i=1}^n I(X_{ki}\le x)$ with $I(\cdot)$ being the indicator function. We can also turn to employ imputation technic that

 \begin{align*}
 F(y)=EI(Y\leq y)
= E\{\delta I(Y\leq y)+\\ (1 -\delta)\frac{\int_{Y^*}^\infty I(Y\leq y)dF(Y\mid X_{k})}{1 - F(Y^*\mid X_{k})}\}
 \end{align*}
to obtain the estimator
\begin{align*}
\widehat{F}_n(y)= 1/n\sum_{i=1}^n\{\delta_i I(Y_i\leq y)+\\ (1- \delta_i)\frac{\int_{Y_i^*}^\infty I(Y\leq y)d\widehat{F}_n(Y\mid X_{ki})}{1- \widehat{F}_n(Y_i^*\mid X_{ki})}\},
\end{align*}
where
$\widehat{F}_n(y\mid x)$ is the estimator of conditional distribution  $F(y\mid x)$ and given by $\widehat{F}_n(y\mid x)\!=\! \sum_{i=1}^n\frac{\delta_iB_{ni}(x)}{\widehat{G}(Y^*_i\mid x)}I(Y^*_i\leq y)$, where $\widehat{G}(y\mid x)$ is the local Kaplan-Meier estimator (He et al. 2014) of $G(y\mid x )={\rm pr}(C\ge y\mid x)$. More specifically,
\begin{small}
\begin{equation}\label{condG}
\widehat{G}(y\mid x)\!=\! \prod\limits_{i=1}^n\Big\{1\!-\! \frac{B_{ni}(x)}{\sum_{j=1}^nI(Y^*_j\ge Y^*_i)B_{nj}(x)}\Big\}^{I(Y^*_i\leq y,\delta_i=0)},
\end{equation}
\end{small}
where $B_{nj}(x)=K(\frac{x-X_{kj}}{h})/\{\sum_{i=1}^nK(\frac{x-X_{ki}}{h})\}$ $(j=1,\ldots,n$), are the Nadaraya-Watson weights, $h$ is the bandwidth and $K(\cdot)$ is a density function. Therefore, the empirical version of $\mathcal{F}(Y^*_i,\delta_i,X_{ki})$ is
\begin{small}
\begin{equation}{\label{zerofest}}
\widehat{\mathcal{F}}_n(Y^*_i,\delta_i, X_{ki})=\delta_i\widehat{F}_n(Y_i^*)+(1-\delta_i)\frac{\int_{Y^*_i}^\infty \widehat{F}_n(y)d\widehat{F}_n(y\mid X_{ki})}{1-\widehat{F}_n(Y^*_i\mid X_{ki})}.
\end{equation}
\end{small}

We propose an
adjusted  Spearman rank correlation screening utility of $\omega_k$ for a censored response  (SRCS$_{\rm cen}$) given by
\begin{equation}\label{SRCSet2}
\widehat{\omega}_k=1/n\sum_{i=1}^n[\widehat{F}_k(X_{ki})\widehat{\mathcal{F}}_n(Y^*_i,\delta_{i},X_{ki}) ]-1/4,
\end{equation}
which is invariant to under any strictly increasing transformation of the response. Therefore, we choose $\widehat{\omega}_k$ as a marginal utility to measure the importance of $X_k$ for response $Y$ in the presence of censoring.
 The corresponding screening set is defined as
\begin{equation}\label{estset}
  \widehat{\mathcal{A}}=\{k: |\widehat{\omega}_k|\ge cn^{-\tau}, 1\leq k\leq p\},
\end{equation}
where $c$ and $\tau$ are pre-determined thresholding values defined by Condition (C1) below.
We note that if all $\delta_i$'s are equal to 1 (i.e., all observed responses are complete), $\widehat{\omega}_k$ in (\ref{SRCSet2}) can be rewritten as $\widehat{\omega}_k=1/n^3\sum_{j=1}^nR_{kj}Q_j-1/4$, where
$R_{kj}=\sum_{i=1}^nI(X_{ki}\le X_{kj})$ and $Q_{j}=\sum_{i=1}^nI(Y_{i}\le Y_{j})$ denote the rank of $X_{kj}$ in all observations of $\boldsymbol{X}_{k}$ and  the rank of $Y_{j}$ in all observations of $\boldsymbol{Y}$, respectively.
 By setting $\bar{R}_{k}=\frac 1n\sum_{j=1}^nR_{kj}$  and $\bar{Q}=\frac 1n\sum_{j=1}^nQ_{j}$, we can express  Spearman's rank-correlation coefficient as
\begin{small}
\begin{align}
\nonumber
\label{srcs}
\rho_{kn}\!=\! \frac{\sum_{j=1}^n(R_{kj}-\bar{R}_{k})(Q_j\!-\! \bar{Q})}{\sqrt{\sum_{j=1}^n(R_{kj}\!-\! \bar{R}_{k})^2
\sum_{j=1}^n(Q_{j}\!-\! \bar{Q})^2}}
\!=\! \\ 12\big\{\frac{1}{n(n^2\!-\! 1)}\sum_{j=1}^nR_{kj}Q_j\!-\! \frac{1}{4}\frac{n\!+\! 1}{n\!-\!1}\big\},
\end{align}
\end{small}
where the detailed derivative process of (\ref{srcs}) is shown in the Appendix. Obviously, the form of $\rho_{kn}$ is analogous to that of $\widehat{\omega}_k$ and  converges to $12\omega_k$.  Therefore, we call this screening procedure Spearman rank correlation screening  with a screening utility $\widehat{\omega}_k$ (SRCS$_{{\rm cen}}$) in (\ref{SRCSet2}).  The Spearman correlation is a nonparametric measure of the statistical dependence between two variables. Unlike Pearson correlation, it assesses how well the relationship between two variables can be described using a monotonic function. This property
allows us to discover the nonlinear relationship between the response and
predictor values. Therefore, it can be directly used to deal with semiparametric models such as those of transformation regression models and single-indices models with monotonic constraints on the link function.

\section{Theoretical Properties}

In this section, we investigate the sure-screening, rank-consistency  and  false-discovery controlling properties of the proposed screening procedures.
Without specifying any regression model of $Y$ and $\boldsymbol{X}=(X_1,\ldots,X_p)^{\T}$,
where $p\gg n$ (with $n$ being the sample size), we define the active predictor subset as
\begin{align}
\nonumber
  \mathcal{D}=\{k:F(y\mid{\bf X})\;{\rm functionally\;depends\;on} \\ \;X_k\;{\rm for\;some}\;y, k=1, \ldots, p\},\nonumber
\end{align}
where $F(y\mid {\bf X})=\mbox{pr}(Y\le y\mid {\bf X})$.
 Then, the sparsity assumption states that $p\gg|\mathcal{D}|$. Our goal is to recover the active set $\mathcal{D}$ as precisely as possible. To this end, we apply the screening procedure depicted in Section 2 for each pair $(X_k,Y)$ as  a marginal utility to measure the importance of $X_k$ for the response $Y$.
We require the following conditions.

(C1)
There exist positive constants $c>0$ and $0\leq\tau< 1/2$ such that $\min_{k\in\mathcal{D}}|\omega_k|>2cn^{-\tau}.$

(C2) $
\min_{k \in \mathcal{D}} |\omega_k| -  \max_{k \notin \mathcal{D}} |\omega_k| \geq c_1n^{-\kappa}$,  where $c_1$ and $\kappa$  are some positive constants.

(C3)
 $Y$ and $C$ are independent given covariates $X_k$ ($k=1, \ldots, p$). $G(y\mid x)$ has uniformly bounded first derivative and bounded (uniformly in $y$) second-order partial derivatives with respect to $x$. Furthermore,  $\inf_x$pr$(y\leq Y_i\leq C_i\mid x)\ge\lambda_1>0$ for some positive constant $\lambda_1$ and any $y\in[0,b_H]$, and $y_1\leq \sup\{y:G(y\mid x)>0\}\leq y_2$ uniformly in $x$ for some positive constants  $y_1$ and $y_2$; The kernel function $K(\cdot)$ is a probability density function
such that it is bounded
and has compact support.

Condition (C1) allows the minimum true signal to be on the order of $n^{-\tau}$, degenerating  to zero as the sample size
increases. Furthermore, this condition can be relaxed by assuming $c= O(n^{-\psi})$ with $0<\psi<2\tau$ \cite{cui2015model}. Under the relaxed
condition, the sure-screening property in Theorem \ref{thm1} still holds, but the convergence rate becomes relatively slower. Condition (C2) ensures that the screening utility can separate informative and non-informative predictors well at the population level, and it is much weaker than the partial-orthogonality condition, i.e., $|\omega_k|\neq 0$ for $k\in\mathcal{D}$ and $|\omega_k|=0$ for $k\notin\mathcal{D}$.
 Condition (C3) is commonly used in the survival analysis literature
to ensure that the Kaplan-Meier estimator and its reciprocal function are well behaved \cite{he2013quantile}.

\begin{theorem}\label{thm1}\quad
  {\rm (i) (Sure-Screening Property)} If Condition (C3) and other conditions in Lemmas hold, there exists a positive constant $b$ depending on $c$, such that
$$ P\left(\max_{1\leq k\leq p}|\widehat{\omega}_k-\omega_k|\geq cn^{-\tau}\right)
\le O[p(n+1)\exp(-bn^{1-2\tau})];$$
and under  Conditions (C1) and (C3),
$$ {\rm pr}(\mathcal{D} \subset \widehat{\mathcal{A}}) \geq 1-O[|\mathcal{D}|(n+1)\exp(-bn^{1-2\tau})].$$
 {\rm (ii) (Rank-Consistency)} If Conditions (C2), (C3), and the additional condition $\log(p)= o(n^{1-2\kappa})$ with $\kappa< 1/2$ hold, then
$$\liminf_{n \rightarrow \infty} ( \min_{k\in \mathcal{D}} |\widehat{\omega}_k| - \max_{k \notin \mathcal{D}} |\widehat{\omega}_k|  )>0.$$
\end{theorem}\quad

\vspace{.1in}
Theorem 1 gives the sure-screening and rank-consistency properties of our proposed screening procedure.
The conditions for the sure-screening property are milder, because we do not require the regression
function of $Y$ onto $\boldsymbol{X}$ to be linear and there are few requirements on the moments of covariates. Since $\omega_k$ inherits the robustness of a distribution function, our proposed screening utility is robust against heavy-tailed distributions of predictors and the presence
of potential outliers.  The exponential tail probability bound of the sure-screening property is less than or equal to that of Wu and Yin \cite{wu2015conditional} and the equality holds only if the quantile $q$ in CQScen($q$) is adopted as 0.5.
We find that our method can handle the NP-dimensionality $\log(p)=O(n^\zeta)$, where $\zeta<1-2\tau$ with $0\leq\tau<1/2$, which depends on the minimum true signal strength. In this case, we have
\begin{align}
\nonumber
   {\rm pr}(\max_{1\leq k\leq p}|\widehat{\omega}_k\!-\! \omega_k|\leq cn^{\!-\! \tau})\ge \\ 1- O[\Lambda\exp\{\!-\! \Lambda n^{1-2\tau}\!+ \log(n\!+\! 1)\}],
\end{align}
where $\Lambda$ is some constant. The rank-consistency property implies that   
the values of $|\widehat{\omega}_k|$ of  active predictors can be ranked ahead inactive ones with high probability. Thus we can separate the active and inactive predictors by taking an ideal threshold value following Mai and Zou \cite{mai2015fused}.

\begin{theorem}\label{thm2}\quad  {\rm (Controlling false discovery)}
 If Condition (C3) holds, there exists a positive constant $b'$ depending on $c$,
$${\rm{pr}}\{|\widehat{\mathcal{A}}|\leq n^{\tau}\sum\limits_k|\omega_k|/c\} \geq 1-O(n)p\exp(-b'n^{1-2\tau}). $$
\end{theorem}\quad
 Theorem \ref{thm2} implies that the model obtained after screening is of polynomial size with high
probability. Although we investigate the theoretical conclusion on controlling the false-positive rate in Theorem \ref{thm2}, the result is conservative for screening purposes because the lower false-positive rate may lead to larger false-negative error. Then an alternative strategy to specify $\widehat{\mathcal{A}}$ practically is to use $\widehat{\mathcal{A}}=\{k: |\widehat{\omega}_k| \;{\rm is\;among\;the\;} d_n{\rm th}\;{\rm largest}\}.$
We note that the sure-screening property of the SRCS$_{\rm cen}$ filter
does not involve $d_n$ or the censoring rate explicitly, leading to tremendous practical convenience for our choice of $d_n$, because we can utilize a
reasonably large $d_n$ to guarantee a high probability of the hold of the sure-screening property.
As Mai and Zou \cite{mai2015fused} suggested we can use $d_n = a[n/\log(n)]$, where
$a$ is some constant whose value may reflect the researchers' prior knowledge of the number
of susceptible predictors, or the budget limitations \cite{song2014censored}. Therefore, its choice is flexible and a more conservative choice could be $d_n\leq n$,
so that a
follow-up regression analysis could be performed in a $p<n$ scenario.

\section{Simulation Studies}
 Simulation studies were conducted to evaluate the performance of the proposed new feature-screening procedure and to compare it with existing screening methods.
We consider six model-free methods, including
 rank correlation screening (RCS) \cite{li2012robust}, censored rank independence screening (${\rm RCS_{cen}}$) \cite{song2014censored}, 
conditional quantile screening with complete and censored responses (CQS($q$), CQScen($q$), where $q$ represents the quantile used) \cite{wu2015conditional}, respectively, our proposed SRCS$_{\rm{cen}}$ in (\ref{SRCSet2}), and the
SRCS defined in (\ref{srcs}). Naturally, we want to know whether we can simply use the screening procedures developed for a complete response in the presence of censoring. To answer this question, we design a naive screening procedure for  RCS, CQS($q$) and SRCS based on datasets $\{(\boldsymbol{X}_i^{\T}, Y^*_i): i=1, \ldots, n\}.$ First, we focus on checking the model-based behaviors of screening procedures combined with some shrinkage methods. Then we aim to at show the robust model-free performance of the proposed nonparametric screening procedure.

\subsection{Linear Model}
In the linear model, we mainly investigate how to decide the true model size.
Considering that some truly unimportant covariates
are also retained in the screening stage, we next perform the
 penalized method to further remove these covariates, and specify the eventual true model size.

\begin{Example}\label{eam0}\quad ({\it  Linear model}).
In this simulation study, we suppose that censored response variable $Y$  takes the following linear model
$$Y_i=X_i^{\T}\boldsymbol{\beta}+\epsilon_i,$$
where $X_i=(x_{i1},\ldots,x_{ip})$ and $(x_{i1},\ldots,x_{i5})$ are derived from Unif(0,1), $(x_{i6},\ldots,x_{ip})$ are assumed to be generated from a multivariate normal distribution with zero mean and covariance matrix $\Phi=(d_{jl})$ with $d_{jl}=0.7^{|j-l|}$, and the censored variable $C_i$ was generated from $C_i=X_i^{\T}\boldsymbol{\theta}+\varepsilon_i$,
 $\epsilon_i$ and $\varepsilon_i$ were assumed to follow the standard normal distribution $\mathcal{N}(1,1)$ and $\mathcal{N}(0,1)$, respectively. We set the sample size to $n=200$ and the number of regressors to
$p=3,000$. The true coefficients $\boldsymbol{\beta}=(\beta_1,\cdots,\beta_p)^{\T}=(3,1.5,0,0,2,\boldsymbol{0}_{p-5})$,
 $\boldsymbol{\theta}=(\theta_0,0,0,-4,-4,\boldsymbol{0}_{p-4})^{\T}$, and $\theta_0$ were chosen to achieve censoring ratios of $45\%$ and $65\%$.
\end{Example}

In this simulated dataset, a total of 200 simulation replications were conducted. Let
$\widehat{\boldsymbol{\beta}}_{(k)}=(\widehat{\beta}_{1(k)},\cdots,\widehat{\beta}_{p(k)})^{\T}$
be the estimator realized in the $k$th simulation replication. Then, the
model selected by
$\widehat{\boldsymbol{\beta}}_{(k)}$ is given by
$\widehat{\mathcal{M}}_{(k)}= \{j:| \widehat{\beta}_{j(k)}| > 0\}$ and
the corresponding average model size (AMS) is $\frac{1}{200}\sum_{k=1}^{200}
|\widehat{\mathcal{M}}_{(k)}|$, where $|\widehat{\mathcal{M}}_{(k)}|$ denotes the cardinality of $\widehat{\mathcal{M}}_{(k)}$.  Let $\mathcal{D}^*=\{j:\beta_j\neq 0\}$  denote the index of the true model; we evaluate the coverage
probability (CP) as $\frac{1}{200}\sum_{k=1}^{200}I(\mathcal{D}^*\subseteq\widehat{\mathcal{M}}_{(k)})$, which measures how
likely it is for all relevant variables to be discovered by one particular
method. To characterize the capability of this method  in
producing sparse solutions, we define the
percentage of correct zeros (PCZ; $\%$) as $\frac{100\%}{p-|\mathcal{D}^*|}\big\{
\frac{1}{200}\sum_{k=1}^{200}\sum_{j=1}^{p}I(\widehat{\beta}_{j(k)}=0)I(\beta_j=0)\big\}$
and percentage of incorrect zeros (PIZ; $\%$) as $\frac{100\%}{|\mathcal{D}^*|}\big\{
\frac{1}{200}\sum_{k=1}^{200}\sum_{j=1}^{p}I(\widehat{\beta}_{j(k)}=0)I(\beta_j\neq0)\big\}$.
If all zero coefficients are correctly identified for all inactive
predictors and no sparse solution is mistakenly estimated for all
relevant variables, the true model is perfectly identified, meaning that giving larger PCZ and smaller PIZ imply a good model-fitting procedure. We utilize the residual-mean-square (RMS) error to examine the performance of the estimated true coefficients, i.e., RMS=$\frac{1}{200}\sum_{k=1}^{200}\sum_{j\in\mathcal{D}^*}\sqrt{\frac{1}{|\mathcal{D}^*|}(\widehat{\beta}_{j(k)}
-\beta_{0j})^2}$. We choose the bandwidth $h=\sigma_{X_k}n^\alpha$, $\alpha=-1/5$, where $\sigma_{X_k}$ is the estimated standard deviation of $X_k$ in
the sample and adopt standard normal distribution as the  kernel density function $K(\cdot).$

To prepare candidate models for fitting, we ordered the regressors via the suggested  SRCS$_{{\rm cen}}$ and the above mentioned screening methods, i.e.,  RCS$_{{\rm cen}}$, CQScen(0.5), SRCS, CQS and RCS.   We separated the first $\lceil n/\log(n)\rceil\approx38$ active predictors using each screening method, and then  considered conducting variable selection and parameter estimation procedures using a combing Buckley-James-type least square objective function and the corresponding penalized functions, i.e.,
\begin{align}
\label{cenls}
\nonumber
\ell_{p}(\boldsymbol{\beta} ; \lambda)=\frac{1}{2}\left\|\boldsymbol{Y}\left(\boldsymbol{\beta}, \widehat{F}_{\boldsymbol{\beta}}\right)-\boldsymbol{X} \boldsymbol{\beta}\right\|^{2}-\\ \frac{1}{2 n}\left[\sum_{i=1}^{n}\left\{Y_{i}\left(\boldsymbol{\beta}, \widehat{F}_{\boldsymbol{\beta}}\right)-X_{i}^{\top} \boldsymbol{\beta}\right\}\right]^{2}+\sum_{j=1}^{p} p_{\lambda}\left(\left|\beta_{j}\right|\right),
\end{align}
where $\boldsymbol{Y}(\boldsymbol{\beta},F)=(Y_1(\boldsymbol{\beta},F),\cdots, Y_n(\boldsymbol{\beta},F))^{\T},$ and
$Y_i(\boldsymbol{\beta},F)=E(Y_i\mid X_i,  Y_i^*, \delta_i)
=\delta_iY_i^*+(1-\delta_i)\Big\{ X_i^{\T}\boldsymbol{\beta}+\frac{\int_{Y_i^*- X_i^{\T}\boldsymbol{\beta}}^\infty tdF_{\boldsymbol{\beta}}(t)}{1-F_{\boldsymbol{\beta}}(Y_i^*- X_i^{\T}\boldsymbol{\beta})}\Big\}$. Here
$\widehat{F}_{\boldsymbol{\beta}}$ is the Kaplan-Meier estimator of $F_{\boldsymbol{\beta}}$ given by
$\widehat{F}_{\boldsymbol{\beta}}(t)=1-\prod_{i:\upsilon_i(\boldsymbol{\beta})\leq t}\left\{1-\frac{1}{G_n(\boldsymbol{\beta},\upsilon_i(\boldsymbol{\beta}))}\right\}^{\delta_i},
$
and $\upsilon_i(\boldsymbol{\beta})=\min(\zeta_i(\boldsymbol{\beta}), \epsilon_i(\boldsymbol{\beta}))$, $\epsilon_i(\boldsymbol{\beta})=Y_i- X_i^{\T}\boldsymbol{\beta}$,
  $\zeta_i(\boldsymbol{\beta})=C_i- X_i^{\T}\boldsymbol{\beta}$, $i=1,\ldots, n$, $G_n(\boldsymbol{\beta},u)=\sum_{i=1}^nI(\upsilon_i(\boldsymbol{\beta})\ge u).$
Next we generate the plug-in response $Y_i(\boldsymbol{\beta},\widehat{F}_{\boldsymbol{\beta}})$. $p_\lambda(\cdot)$ are some sparsity-inducing penalties and here we consider SCAD, MCP and LASSO to obtain penalized estimator; the optimal value of $\lambda$ can be selected via the following EBIC
criterion  \cite{chen2008extended, wang2009forward},
\begin{small}
\begin{align*}
{\rm EBIC}(\mathcal{M})=\log n^{-1}\boldsymbol{Y}^{\T}(\boldsymbol{\beta}_{\mathcal{M}},
\widehat{F}_{\boldsymbol{\beta}_{\mathcal{M}}}) [I_n-\boldsymbol{X}_{\mathcal{M}}
(\boldsymbol{X}_{\mathcal{M}}^{\T}\boldsymbol{X}_{\mathcal{M}})^{-1}\\
\boldsymbol{X}_{\mathcal{M}}^{\T}] \boldsymbol{Y}(\boldsymbol{\beta}_{\mathcal{M}},
\widehat{F}_{\boldsymbol{\beta}_{\mathcal{M}}})
+n^{-1}
|\mathcal{M}|(\log n+2\log p),
\end{align*}
\end{small}
where $\boldsymbol{\beta}_{\mathcal{M}}$ and $\boldsymbol{X}_{\mathcal{M}}$ are respectively, the sub-parameter  coefficients and  sub-design matrix corresponding to $\mathcal{M}$.  To calculate all slopes, we specify a proper initial value and adopt a Buckley-James iterative procedure \cite{miller1982regression} to update the
the regression estimators until input of some convergence criterion.

\begin{small}
\begin{table}
\centering
\caption{Simulation results by different screening procedures and variable selection methods in Example \ref{eam0}.}
\scriptsize
\scalebox{0.55}{
 \begin{tabular}{ccccccccccccc}\hline \hline
&&\multicolumn{5}{c}{CR=45\%} &&\multicolumn{5}{c}{CR=65\%}\\
\cline{3-7} \cline{9-13}
Screening &Selection& & & && && &  & & & \\
method &method& CP(\%)&  PCZ(\%)&PIZ(\%)&RMS &AMS&&  CP(\%)& PCZ(\%)&PIZ(\%)&RMS &AMS \\ \hline
SRCS$_{\rm cen}$   &SCAD&91.50&99.13&2.50 &0.49 &3.21 &&86.50&98.57&5.67&0.82&2.42 \\
       &MCP&91.50 &	99.16&2.50 &0.49 & 3.24&&86.50&98.53&5.67&0.84 & 2.43 \\
        &LASSO&91.50 &99.59&2.50 &0.86 & 3.01&&86.00&98.99& 2.74&1.49&2.23  \\
CQS(0.5)$_{\rm cen}$   &SCAD&63.50&98.87& 14.65 & 0.66&1.72&&60.00&98.42&15.22&0.91 & 1.21 \\
       &MCP&63.50&	98.99&14.65&0.66&1.72 &&60.00&98.43&15.22&0.91&1.20 \\
        &LASSO&63.50 &99.47&15.33 &0.99&1.61&&58.50&99.39&14.02& 1.55&1.18  \\
RCS$_{\rm cen}$   &SCAD&81.50&99.11&11.27 &0.68&1.91&&11.00&98.78&45.07&1.35  & 0.11 \\
       &MCP&81.50&	99.30&11.27&0.68&1.92 &&11.00&98.79&45.07  &1.35& 0.11 \\
        &LASSO&81.50&	99.79&11.27&1.09& 0.64&&10.00&99.69&43.21  &1.68&0.00  \\
SRCS   &SCAD&65.50&99.10&15.20&0.94&0.72&&22.50&98.62&33.43& 1.16&0.25 \\
       &MCP&65.50&	99.23&15.20&0.95 & 0.72&&22.50&98.64&33.43&1.16  & 0.25 \\
        &LASSO&65.50&99.48&15.20&1.23 & 0.02&&20.50&99.34&39.54&1.69 &0.00  \\
CQS(0.5)  &SCAD&31.00&98.95& 33.43 & 0.95&0.12 &&4.50&98.62&59.08& 1.78& 0.00 \\
       &MCP&31.00&	99.15& 33.43&0.95&0.12&&4.50&98.69&59.08&1.78 &0.00  \\
        &LASSO&31.00&99.84& 35.62 &1.56 &0.00&&4.00&99.92&61.27& 1.99 &0.00  \\
RCS  &SCAD&54.50&98.86& 23.50&0.85 &0.92&&14.00&98.55&40.22&1.36& 0.06 \\
       &MCP&54.50&	98.83& 23.50&0.85 &0.93 &&14.00&98.65&40.22  &1.36& 0.06 \\
        &LASSO&54.00&	99.55& 26.52&1.44 & 0.02&&13.50&99.26&43.27  &1.98&0.00  \\
\hline
\end{tabular}}
 \\
\end{table}
\end{small}

Table 1 presents the simulated results for Example \ref{eam0}. We note that (i) traditional screening methods, (i.e., SRCS, CQS, and RCS), fail to recover the true model  combined with the penalized procedure, because according to Table 1, they show lower coverage probability and their estimated average model size approximates zero under the censored case; (ii)
 The SRCS$_{{\rm cen}}$+$penalized$ $method$  shows the best performance in terms of the highest coverage probability; (iii) although all methods show higher percentages of correct zeros, the SRCS$_{{\rm cen}}$+$penalized$ $method$  shows a lower percentage of incorrect zeros, this result is attributed to the wrongly selected predictors in the first screening step; (iv) under the SRCS$_{{\rm cen}}$ screening method, the two penalization SCAD (i.e., SRCS$_{{\rm cen}}$-SCAD) and
MCP (i.e., SRCS$_{{\rm cen}}$-MCP)  procedures have similar model fitting performance in terms of nearly equal  RMS errors. Additionally, SRCS$_{{\rm cen}}$-concave penalty procedures generate much smaller RMS errors than SRCS$_{{\rm cen}}$-Lasso; (v) Unlike the estimated result of RCS$_{{\rm cen}}$,  the coverage probability obtained by
SRCScen still approximates one under a 65\% censoring rate, implying that our proposed method remains robust, whereas RCS$_{{\rm cen}}$ is dramatically reduced under a 65\% censoring rate; (vi) the SRCS$_{{\rm cen}}$-concave penalties method is superior in recovering the true model, because according to Table 1, the estimated average model size under this method approximates the true model size, i.e., 3.

\subsection{Other Models}
 To measure the model-free performances of all screening methods, we consider measurements including
(i) the selected model size, i.e., the average number of active variables
contained in the top $m$ selected variables. We consider $m=4,10,$ (i.e., $m_4$ and $m_{10}$) for Example \ref{eam1} and $m=3,10$, $m=2,10$ for Examples  \ref{eam2} and  \ref{eam3} respectively, because $m_2$, $m_3$ and $m_{10}$ have corresponding numbers of true active predictors given by 4, 3 and 2, respectively;
(ii) the coverage probability ($\mathcal{P}_{{\rm all}}$) of the the top $\lceil n/\log(n)\rceil$ estimated  variables covering the true ones; (iii)
 the median minimum model size, i.e., the minimum number of predictors needed to keep all the active predictors; (iv) the interquantile range (IQR) of the minimum model size needed to include all active predictors. Here we adopt the range of the 0.8 and 0.2-quantiles.
 To show the robustness of
our proposed nonparametric screening procedures, we consider the following three models. For each setting, we consider $n=200$, $p=3000$ and  200 replicates.

\begin{Example}\label{eam1}\quad
({\it  Linear transformation model}). We adopt the model setting of \cite{song2014censored} and generate failure time $T_i$ from the class of linear transformation models
$$H(T_i )=\boldsymbol{X}_i^{\T}\boldsymbol{\beta}+\epsilon_i,\;i=1,\cdots,n,$$
where $Y_i=H(T_i)$ and $H(t)=\log{0.5(e^{2t}-1)}$, $\boldsymbol{\beta}=(-1,-0.9,\boldsymbol{0}_4^{\T},0.8,1,\boldsymbol{0}_{p-8}^{\T}),$ $\boldsymbol{X}_i\sim N(0,\Sigma)$ and $\Sigma=(\sigma_{kj})_{1000\times1000}$ with $\sigma_{kk}=1$ and $\sigma_{kj}=0.5^{|k-j|}$.  Although $Y_i$ is unobserved for an unknown function $H(\cdot$), our method is still applicable to this model setting for its invariant under any strictly increasing transformation. The model is flexible, because if  we adopted three error distributions including the standard normal distribution, and the standard logistic distribution or standard extreme-value distribution, this model
 would correspond to a normal transformation model,  a proportional-odds model, or a proportional-hazards model, respectively. The censoring time was generated from Unif(0, $\kappa$) or  Unif(0, $\kappa\exp(|X_1-X_2|))$ for random and non-random censoring, where $\kappa$ was chosen to achieve censoring rates of $45\%$ and $65\%$. To study the robustness of our method, we contaminated the predictors by adding
outliers to them using the strategy of \cite{song2014censored}.
\end{Example}

\begin{Example}\label{eam2}\quad
({\it Additive model}). We consider the additive model of the form: $Y=\sum\limits_{j=1}^pg_j(X_j)+\epsilon$. Set $g_1(X_1)=4X_1,$ $g_2(X_2)=2\tan(\pi X_2/2)$, $g_3(X_3)=5X_3^2$. Assume that the $X_j$ were generated from Unif(0,1) independently and that $\epsilon\sim {N}(0,1)$ is independent of $\boldsymbol{X}$.  The censoring variable was generated from a 2-component normal-mixture distribution plus a constant, i.e.,
$\kappa_1{N}(-5,2)+\kappa_2{N}(5,1)+\kappa_3$ for censoring rates of 45$\%$  and 65$\%$, respectively.
\end{Example}

\begin{Example}\label{eam3}\quad
({\it Single-index-regression model}). The model takes  $Y=(X_1+X_2+1)^3+\epsilon$. The $X_j$ follow the Cauchy distribution independently, and $\epsilon\sim {N}(0,1)$ is independent of $\boldsymbol{X}$.
      We still adopt the strategy of Example \ref{eam2} to obtain the censored variable and achieve censoring rates of 45$\%$  and 65$\%$.
\end{Example}

\begin{small}
\begin{table}
\centering
\caption{Simulation results by different screening methods for Example \ref{eam1} with random censoring.}
\scriptsize
\scalebox{0.47}{
 \begin{tabular}{ccccccccccccc}\hline \hline
&&\multicolumn{5}{c}{CR=45\%} &&\multicolumn{5}{c}{CR=65\%}\\
\cline{3-7} \cline{9-13}
Error&Method& $m_{4}$& $m_{10}$ & $\mathcal{P}_{all}$& Median&IQR && $m_{4}$& $m_{10}$ & $\mathcal{P}_{all}$& Median&IQR \\ \hline
Normal&SRCS$_{\rm cen}$&3.78 &4.00& 1 & 4&0 &&3.35&3.59&0.92&3.90  & 0.25 \\
&CQS(0.5)$_{\rm cen}$&3.72 &3.90& 0.98 &4 &0 &&3.25&3.41&0.86&3.75  & 1.50 \\
       & RCS$_{\rm cen}$& 3.64 & 3.81& 0.92&4& 1.68&& 0.72 & 1.44  & 0.00  &1000&10.2\\
       & SRCS&2.64  &3.20 & 0.63&   12.5 &  72.4      && 0.31  & 0.56  &0.00   &653.5&141.2\\
       &CQS(0.5)&2.35  & 2.97&0.67 &   20 &  50.8     &&  0.22 & 0.44  &  0.00 &458&381.2\\
       & RCS& 3.20 &3.55 & 0.71&   5 &  23.4      && 0.53  &  0.79 &0.00   &452&240.5\\
Logistic&SRCS$_{\rm cen}$&3.75 &3.79&1 & 4&0.15 &&3.51&3.65&0.97&3.90&3.05\\
  &CQS(0.5)$_{\rm cen}$&3.52&3.64& 0.95 & 3.90& 3.42&&3.58&3.60&0.88& 3.75 & 8.25 \\
       & RCS$_{\rm cen}$&  2.85 & 3.20& 0.70&10.5 &  59.4&&0.81& 1.94  &0&1000&0\\
           & SRCS& 1.55 &2.20 & 0.19&  103.5 & 413.4 &&0.19&  0.75 &0&664.5&571.2\\
       &CQS(0.5)& 1.35 &2.00 &0.16 &  119.5&  403.2     &&0.09&0.59&0&538&679.8\\
       & RCS& 2.29 &2.62 & 0.40&  68.5 & 133   &&0.21&1.06&0&478&632\\
Weibull&SRCS$_{\rm cen}$&3.87 &	3.95& 1&4 & 0.15&&3.77&3.90&0.98&4&  1.08\\
         &CQS(0.5)$_{\rm cen}$&3.68 &3.80&0.98 &4 &1.98 &&3.57&3.75&0.90&4&5.24\\
       & RCS$_{\rm cen}$&3.45&3.47&0.85&4&9.2&&2.21&2.32&0.17&197.5&788.2\\
        & SRCS&2.09&2.27 & 0.37& 93&253.4&&1.31&1.39&0.15&344&515.2\\
       &CQS(0.5)& 1.65 &2.38 & 0.35&91&273.4&&1.19&1.76&0.11&215.5&413\\
       & RCS&2.72& 2.72&0.68 &82.5 &133.2&&1.79&2.58&0.29&145.5&234.4\\
\hline
\end{tabular}}
 \\
\end{table}
\end{small}

The simulation result for Example \ref{eam1} is reported in Table 2 and show that the performance of the proposed method is comparable to or better than other screening methods  under random censoring; we note that (i) SRCS$_{{\rm cen}}$  performs better than RCS$_{{\rm cen}}$, because according to Table 2, SRCS$_{{\rm cen}}$ shows much higher coverage probability of the
top  $\lceil n/\log(n)\rceil$ estimated variables covering the true ones,
a larger average number of active variables contained in the top $m_4$ and $m_{10}$, and
smaller 0.8 and 0.2-quantiles ranges of the
minimum model size. However, under a high censoring ratio, RCS$_{{\rm cen}}$ is invalid for screening active predictors, because coverage probability equals zero in
Table 2. (ii) SRCS$_{{\rm cen}}$ generates the same median
minimum model size as CQScen(0.5) in Table 2, but SRCS$_{{\rm cen}}$ outperforms  CQScen(0.5), because SRCS$_{{\rm cen}}$ always shows smaller 0.8- and 0.2-quantile ranges of
minimum model size and larger coverage probabilities;
For simulated results of the additive model in Example \ref{eam2} and the single-index-regression model in Example \ref{eam3}, one can see that our proposed method  SRCS$_{{\rm cen}}$ still performs best among all screening methods because according to Table 4, SRCS$_{{\rm cen}}$ shows a higher coverage probability, larger selected model size and smaller interquantile range than other screening procedures and RCS$_{{\rm cen}}$ still cannot work under a censoring ratio of $65\%$, because the coverage probability approximates or equals zero in Table 4. The screening results of  Examples \ref{eam2} and \ref{eam3} again prove that our proposed nonparametric screening method is a robust model-free procedure.

\begin{table}
\centering
\caption{Simulation results of  different screening methods for Examples \ref{eam2} and \ref{eam3}.}
\scriptsize
\scalebox{0.47}{
 \begin{tabular}{ccccccccccccc}\hline \hline
&&\multicolumn{5}{c}{CR=45\%} &&\multicolumn{5}{c}{CR=65\%}\\
\cline{3-7} \cline{9-13}
Example &Method& $m_{3}$/$m_{2}$& $m_{10}$ & $\mathcal{P}_{all}$& Median&IQR && $m_{3}$/$m_{2}$& $m_{10}$ & $\mathcal{P}_{all}$& Median&IQR \\ \hline
\ref{eam2}&SRCS$_{\rm cen}$&2.51&	2.82& 0.97&3 & 2&&2.36&2.74&0.90& 3 &5.02  \\
&CQS(0.5)$_{\rm cen}$& 2.31&2.62&0.90 &4 & 10.05&&2.05&2.33&0.83& 7 & 15.2 \\
       & RCS$_{\rm cen}$& 1.49 &1.89 &0.41 &  67 &  318      && 0.52  &0.72   &   0.07&435&455.6\\
        & SRCS&2.03  &2.17 & 0.82&   4.5 & 18.4  && 1.03  & 1.69  &0.41   &587&790.6\\
       &CQS(0.5)& 1.42 & 2.01&0.64 &  57.5 & 92.6&& 0.55  & 1.05  & 0.18  &352.5&303.2\\
       & RCS&1.88  & 2.22& 0.79&  6 &47.4&&1.09   & 1.35  & 0.38  &182&539.4\\
\ref{eam3}&SRCS$_{\rm cen}$&1.99 &2&1&2 & 0&&1.85&1.97&0.95& 2 & 0.85 \\
  &CQS(0.5)$_{\rm cen}$&1.98 &2&1 &2 &0 &&1.92&1.96&0.92& 3 & 5.33 \\
       & RCS$_{\rm cen}$&0.51&0.43 &0.31 &520.5 &689  && 0  & 0  &  0 &1000&0\\
       & SRCS&1.37  &1.59 &0.85 & 4 &2.45&& 0.16  &0.24   & 0.00  &445.5&761\\
       &CQS(0.5)& 1.81 & 1.96&0.93 & 4 & 3.82&& 0.07  & 0.18  & 0.00  &626&712.2\\
       & RCS&1.72  &1.79 & 0.95&4 & 2.92&& 0.21  & 0.32  & 0.01  &344&787.8\\
\hline
\end{tabular}}
\\
\end{table}

In summary, the proposed new screening method is more robust than existing ones and provides superior performance. In particular, we can conclude from our extensive simulation studies that applying screening procedures designed for a complete response to a censored response may not be valid.  In addition,
 when a response is censored at a high rate such as 65\%, our screening procedure still works well. To intuitionally expose the mechanism of robustness under a high censoring ratio,
 we assume that ($X_j, Y$) follows a bivariate normal distribution with zero mean vector and
correlation $\rho_j $= Cor($X_j$, $Y$); thus, $X_j$ and $Y$ have standard normal distributions.  We generated a censored variable $C$ from $\mathcal{N}(\eta_0, 0.3^2)$ and chose different values of $\eta_0$  to achieve four censoring ratios (CRs) of 30$\%$, 50$\%$, 70$\%$ and 90$\%$. Thus, the values of screening statistics are the functions of $\rho_j$ and CR. Figure 1 presents the change of the screening utilities of RCScen, CQScen and SRCScen against $\rho_j$ under different CR's. Examination of Figure 1 shows that (i) the three screening utilities are strictly increasing functions of $|\rho_j|$ under relatively low censoring ratios, such as 30$\%$ or 50$\%$;
(ii) such monotonic properties of RCScen disappear under censoring ratios of 70$\%$ and 90$\%$, implying that
RCScen is invalid for screening active predictors under higher censoring ratios; and (iii)
the statistical value of SRCScen increases more quickly with increasing $|\rho_j|$ than does CQScen, especially under higher censoring ratios like 90$\%$. This elucidates that
 SRCScen is more powerful for ranking the importance of covariates.

\section{An Application}
In this section, we applied our proposed screening procedure with a censored response to the mantle cell lymphoma microarray data, available from http://llmpp.
nih.gov/MCL/. The data consist of the survival times of 92 patients and the gene-expression measurements of 8,810 genes for each patient. However, we only considered 6,312 genes for each patient after deleting 2,498 that  appeared to be missing. During the follow-up, 64 patients died of mantle-cell lymphoma and the other 28
were censored, causing a censoring rate of $36\%$. Our goal  is to identify genes with
great influence on patient survival risk. Under a given model size $\lceil 92/\log(92)\rceil=20$, we ranked the most influential of
the 6312 genes using the proposed screening procedure in Section 2 and  two other methods developed by Song et al. \cite{song2014censored} and Wu and Yin \cite{wu2015conditional}.
We summarized the top 20 selected
genes in Table 5 and found 12 genes commonly selected by the three screening
methods. Their unique identifications are 28990, 30157, 27095, 34771, 28346, 28872, 34790, 30334, 25234, 31420, 17326 and 17123. Wu and Yin \cite{wu2015conditional} have confirmed that these genes may be strongly
associated with patients' survival risk. In addition, Gene 28990, i.e., cell division cycle
2, G1 to S and G2 to M, was ranked first by our proposed method and its   importance has also been  identified by Huang and Ma \cite{huang2010variable} and Wu and Yin \cite{wu2015conditional}. Among the top 20 genes, the numbers commonly selected by SRCS$_{{\rm cen}}$ and CQScen(0.5) are 14 with 16 being commonly selected by SRCS$_{{\rm cen}}$ and
RCS$_{{\rm cen}}$, and 12 by CQScen(0.5) and RCS$_{{\rm cen}}$. Therefore, our approach shows the highest total number of common genes with other two methods, i.e., 30.

\begin{table}
\centering
\caption{Top 20 selected predictors for mantle cell lymphoma microarray dataset.}
\scriptsize
\scalebox{0.6}{
 \begin{tabular}{ccccccccccccc}\hline \hline
 order   &${\rm SRCS_{cen}}$&${\rm CQS(0.5)_{cen}}$&${\rm RCS_{cen}}$ \\
1&28990 & 30157&27095\\
2&30157 &28990 &34771\\
3&24418 &25234 &30157\\
4&27095 &34790 &28990\\
5&34771 &31420 &28872\\
6&28346 &28346 &27762\\
7&32187 &17326 &32699\\
8&28872 &24794 &30282\\
9&34790 &26950 &27049\\
10&30334 &27095 &28346\\
11&25234 &28872 &32187\\
12&27678 &34771 &16587\\
13&31420 & 16787&17343\\
14&27762 &30334 &27019\\
15&27049 &24610 &17326\\
16&17326 &17198 &17123\\
17&17343 & 24723&34790\\
18&17198 &26933 &30334\\
19&16787 &17123 &25234\\
20&17123 & 16528&31420\\
\hline
\end{tabular}}
\\
\end{table}

Next, we examine how a variable-screening procedure helps to predict the conditional mean of censored response variables. To this end, we regard the logarithm of patients' survival time as response $Y$ and 6,312  genes as covariates $\boldsymbol{X}=(X_1,\cdots,X_{6132})^{\T}$. All covariates were standardized before applying all methods. Then, we consider the linear model
$$Y_i=\boldsymbol{X}_i^{\T}\boldsymbol{\beta}+\epsilon_i, i=1,\cdots, 92,$$
we require the following two steps to be implemented for model-fitting, i.e., the $Screening$+$penalized$ $methods$ procedure.
The first step is to order the set of $6312$ regressors and adopt the top $\lceil \frac{92}{\log(92)}\rceil=20$ predictors for the six screening index magnitudes (i.e., SRCS, CQS(0.5), RCS, SRCS$_{{\rm cen}}$, CQScen(0.5), and RCS$_{{\rm cen}}$). In the second step we consider combining the Buckley-James-type least-squares objective function and the corresponding penalized functions (i.e., SCAD, MCP, Lasso), i.e., Equation (\ref{cenls}), to obtain a penalized estimator $\widehat{\boldsymbol{\beta}}$; then, we obtain a conditional mean estimator $\widehat{\mu}_i$.
To evaluate the prediction performance of various methods, we adopt bootstrap strategy  and let $B=\{i: \mbox{observation} \; i \; \mbox{is resampled}\}$ as the index set of observations resampled and $\mathcal{N}_i$ denotes the resampled number of $i$. We used the average squared prediction errors (ASPE)
$${\rm ASPE}=\frac{1}{\sum_{i \in B }\mathcal{N}_i\delta_i}\sum_{i \in B }\mathcal{N}_i\delta_i(Y_{i}-\widehat\mu_i)^2.$$
Figures 2 and 3 present boxplots of the ASPEs after 500 replications.
Figure 2 shows that the traditional screening methods (i.e., SRCS, CQS(0.5) and RCS), developed for a complete response perform poorly for a censored response in terms of larger ASPEs compared with that shown in Figure 3.
Examination of Figure 3 shows that
our proposed SRCS$_{{\rm cen}}$ method together with any shrinkage procedure offers a better performance than other methods
in terms of the smallest  median of 500 ASPE values among the three censored screening methods.

\begin{figure}[h]
\begin{subfigure}{.3\linewidth}
  \centering
  \includegraphics[width=1\linewidth]{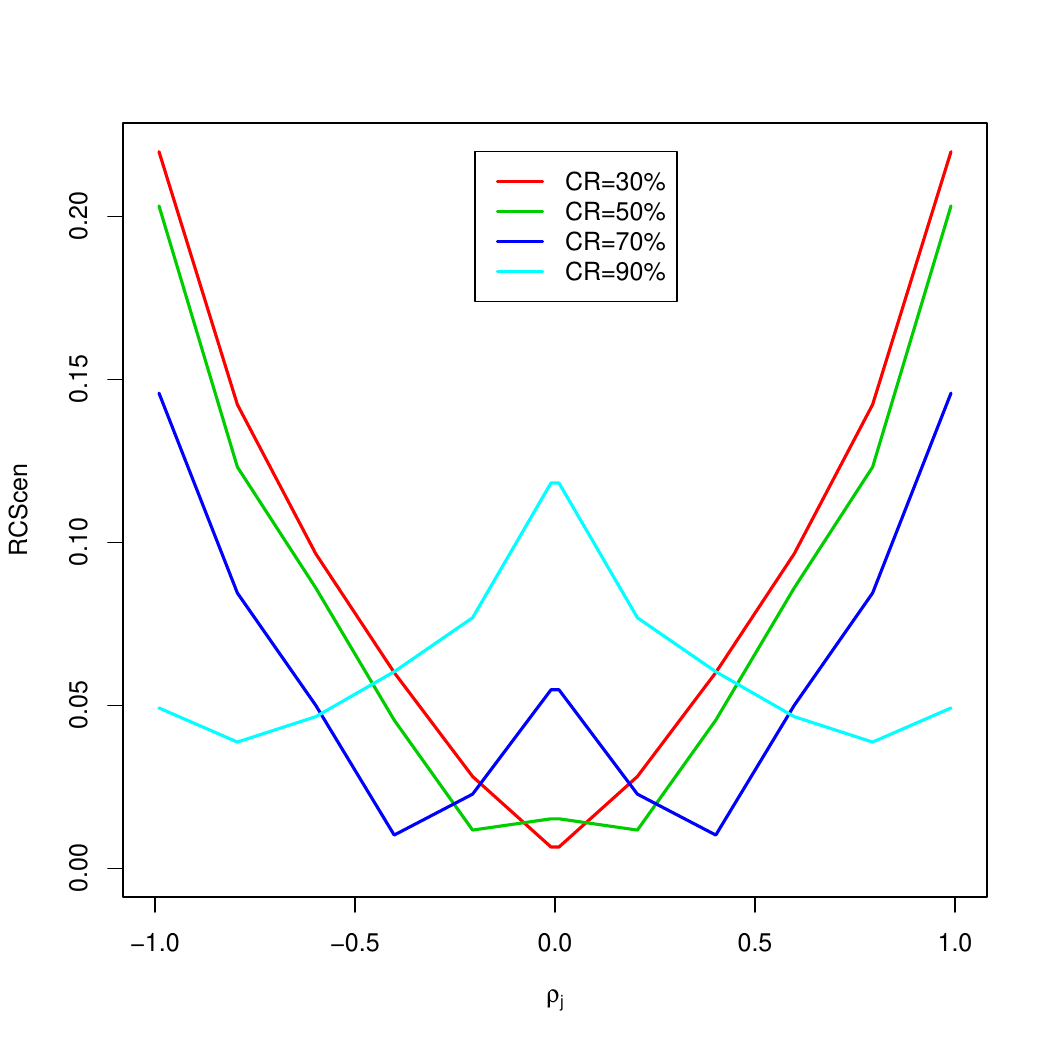}
  \caption{RCScen}
  \label{fig101}
\end{subfigure}%
 \hfill
\begin{subfigure}{.3\linewidth}
  \centering
  \includegraphics[width=1\linewidth]{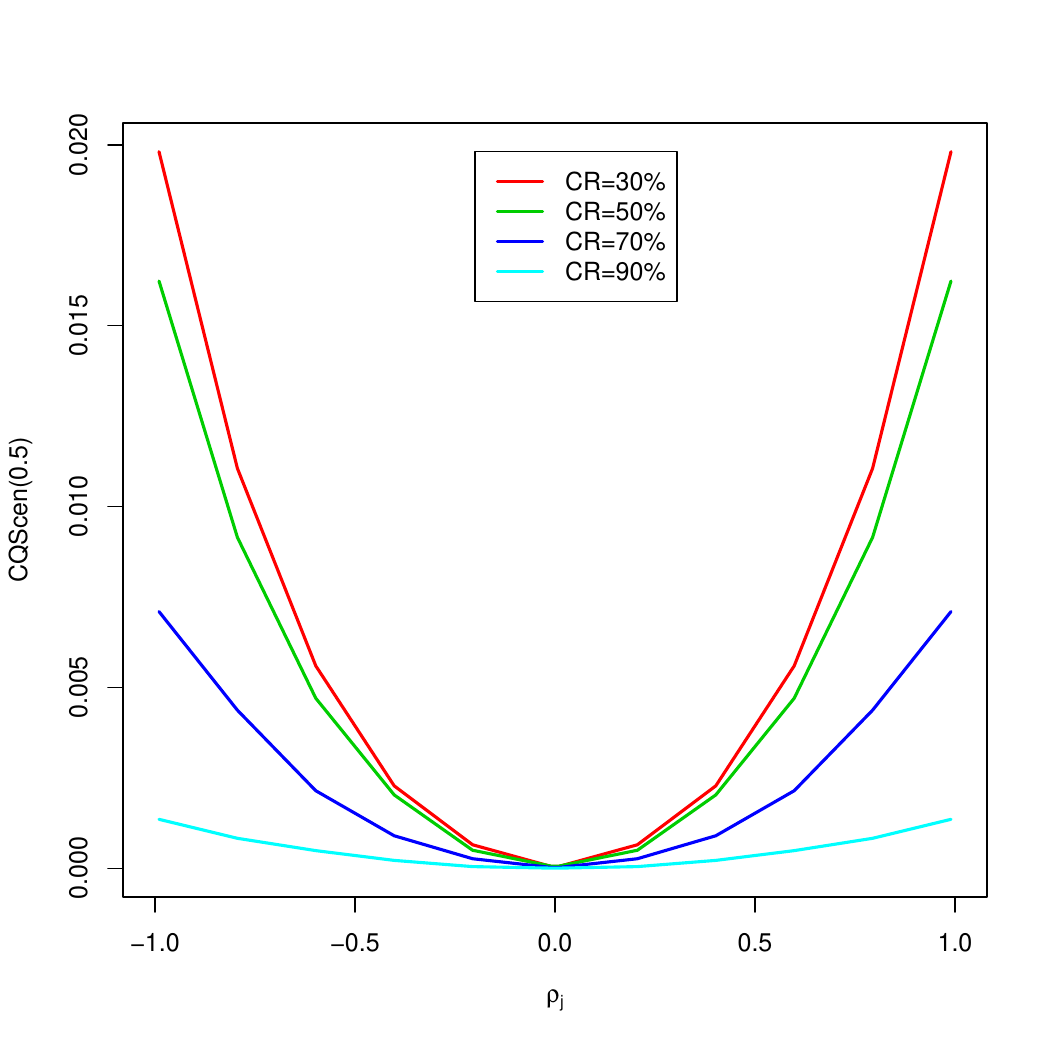}
  \caption{CQScen}
  \label{fig102}
\end{subfigure}
 \hfill
\begin{subfigure}{.3\linewidth}
  \centering
  \includegraphics[width=1\linewidth]{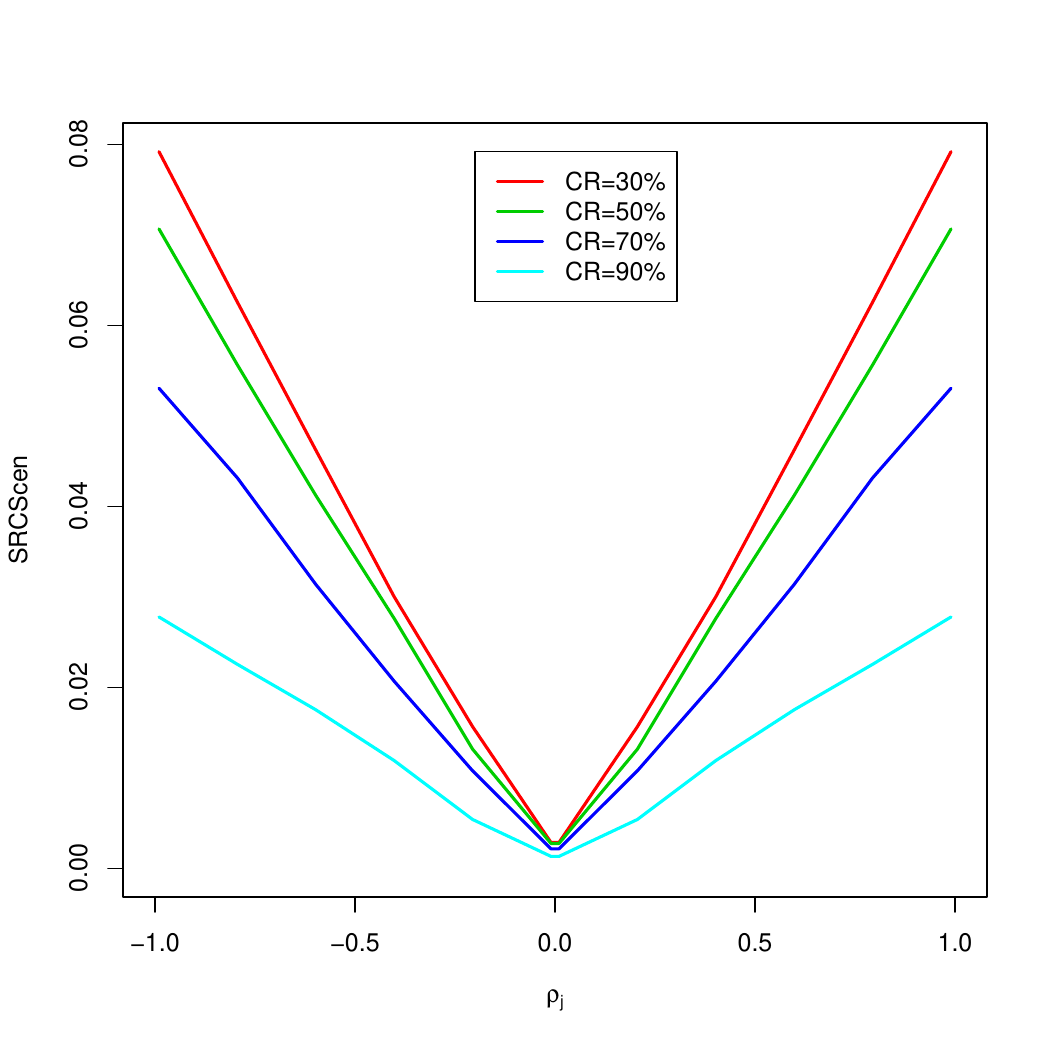}
  \caption{SRCScen}
  \label{fig103}
\end{subfigure}
\caption{
Plots of values of screening statistic. 
}
\label{figg10}
\end{figure}

\begin{figure}[h]
\begin{subfigure}{.3\linewidth}
  \centering
  \includegraphics[width=1\linewidth]{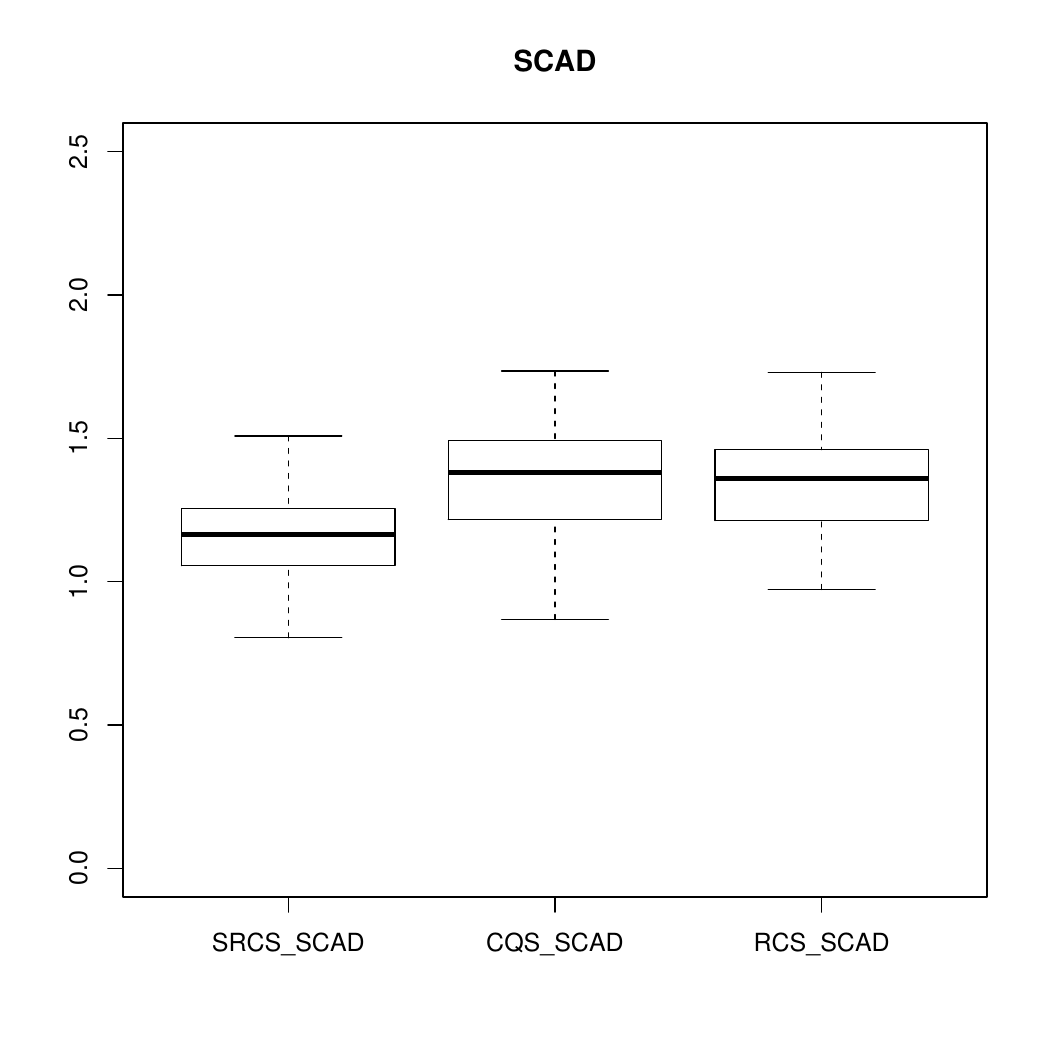}
  \caption{SCAD}
  \label{figb1}
\end{subfigure}%
 \hfill
\begin{subfigure}{.3\linewidth}
  \centering
  \includegraphics[width=1\linewidth]{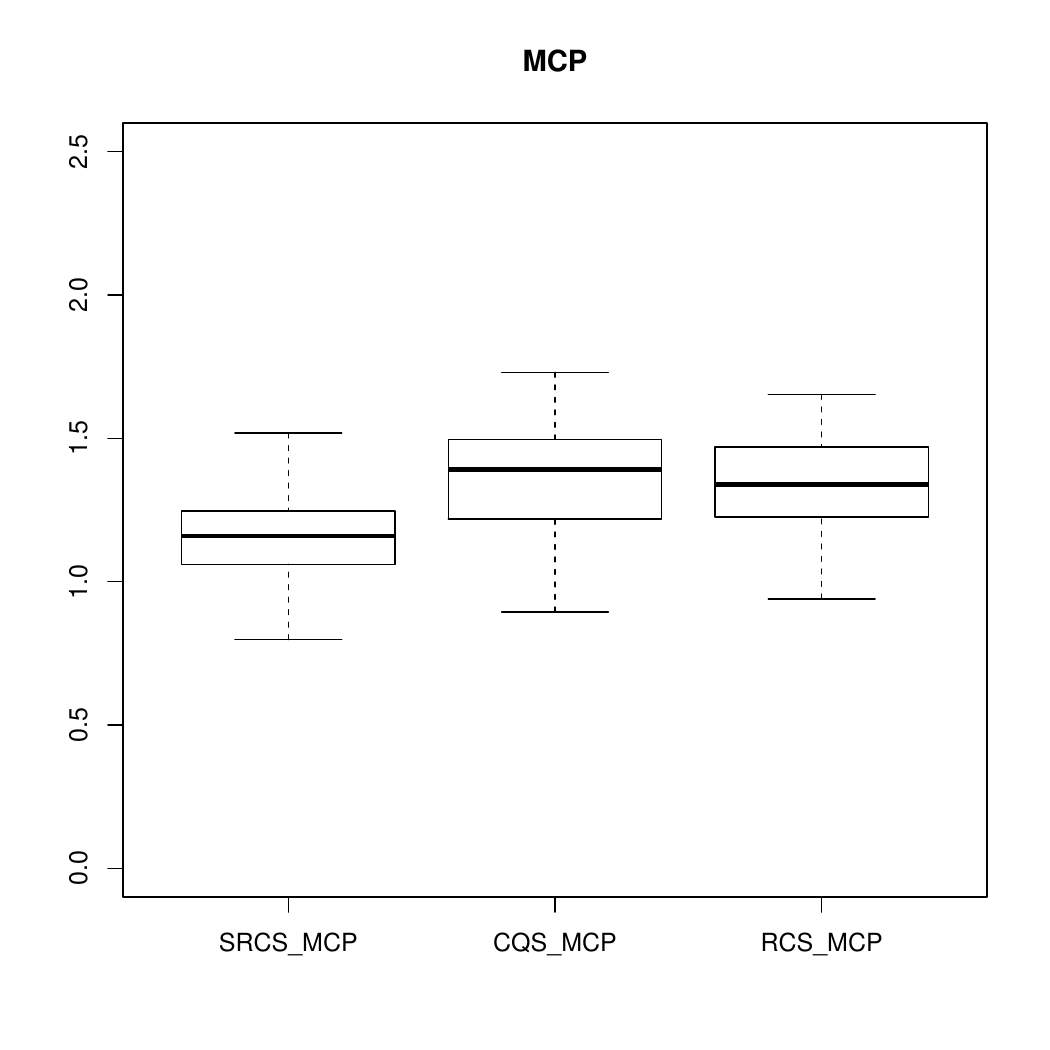}
  \caption{MCP}
  \label{figb2}
\end{subfigure}
 \hfill
\begin{subfigure}{.3\linewidth}
  \centering
  \includegraphics[width=1\linewidth]{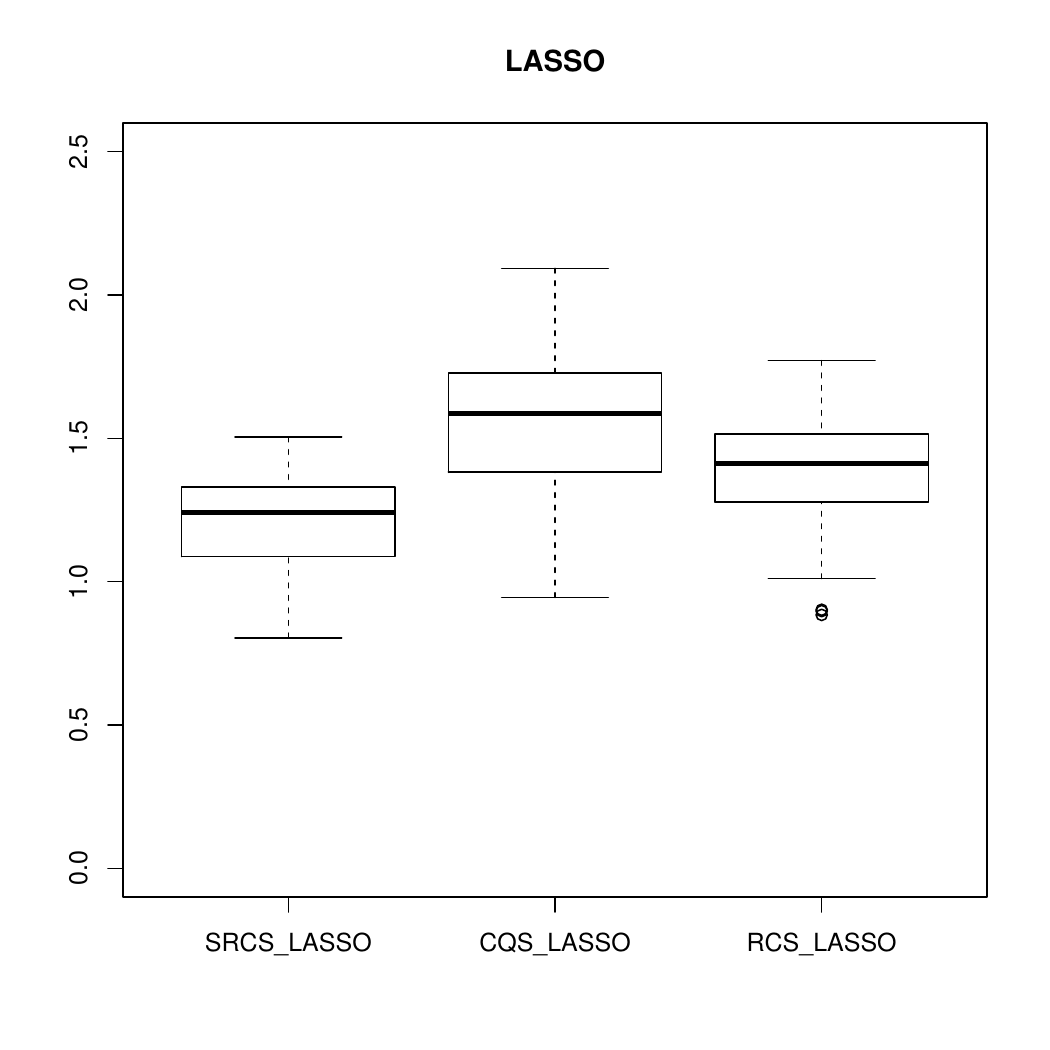}
  \caption{LASSO}
  \label{figb3}
\end{subfigure}
\caption{
Boxplots of averaged squared prediction errors. 
}
\label{fig10}
\end{figure}

\begin{figure}[h]
\centering
\begin{subfigure}{.3\linewidth}
  \centering
  \includegraphics[width=1\linewidth]{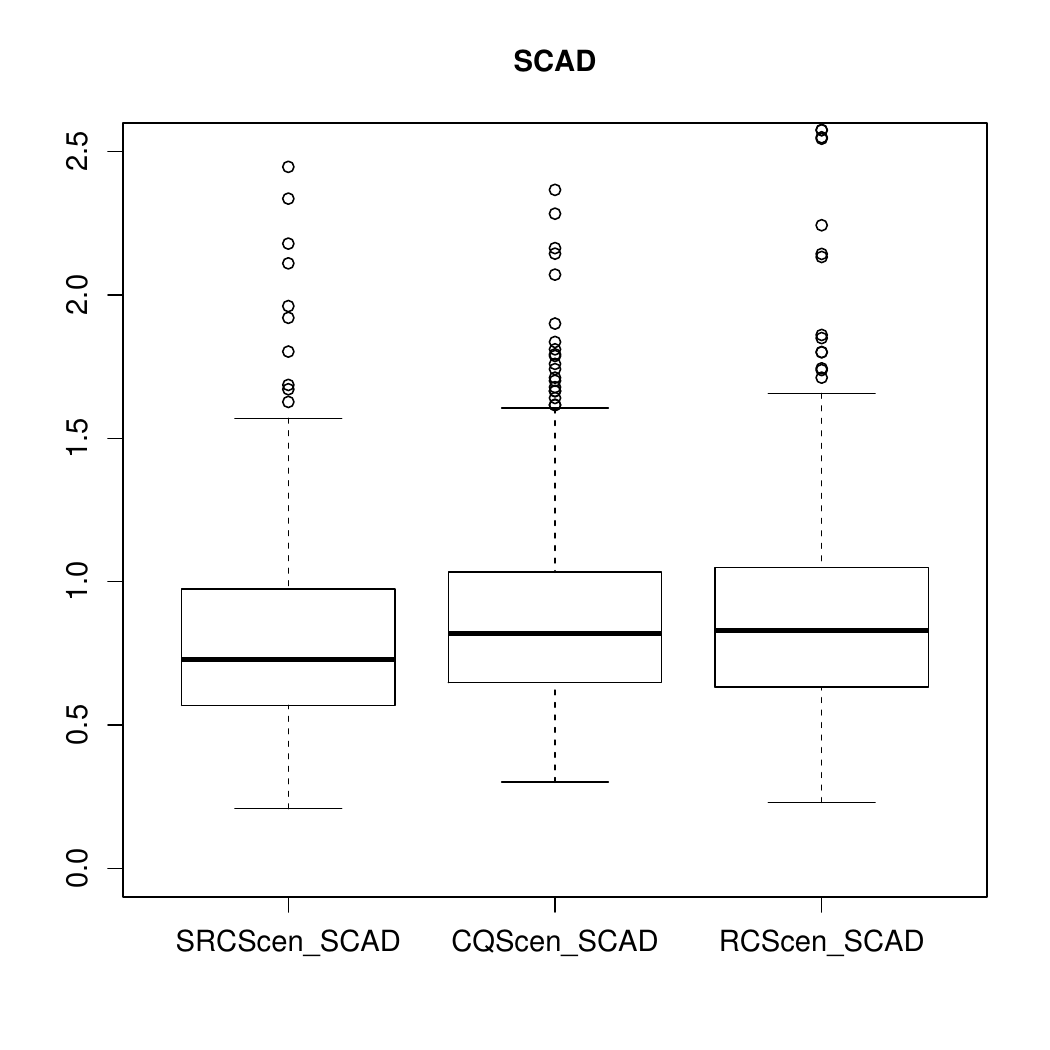}
  \caption{SCAD}
  \label{fig101}
\end{subfigure}%
 \hfill
\begin{subfigure}{.3\linewidth}
  \centering
  \includegraphics[width=1\linewidth]{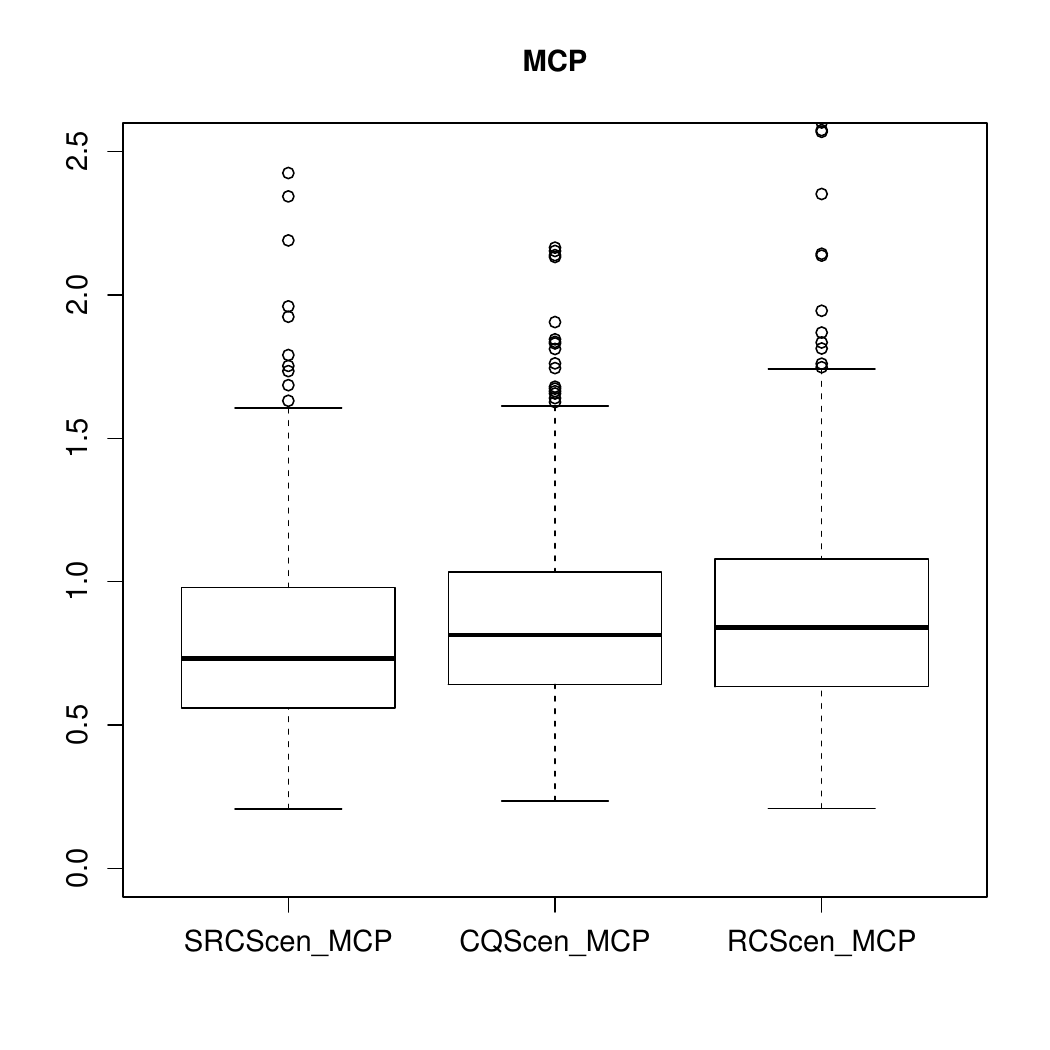}
  \caption{MCP}
  \label{fig102}
\end{subfigure}
 \hfill
\begin{subfigure}{.3\linewidth}
  \centering
  \includegraphics[width=1\linewidth]{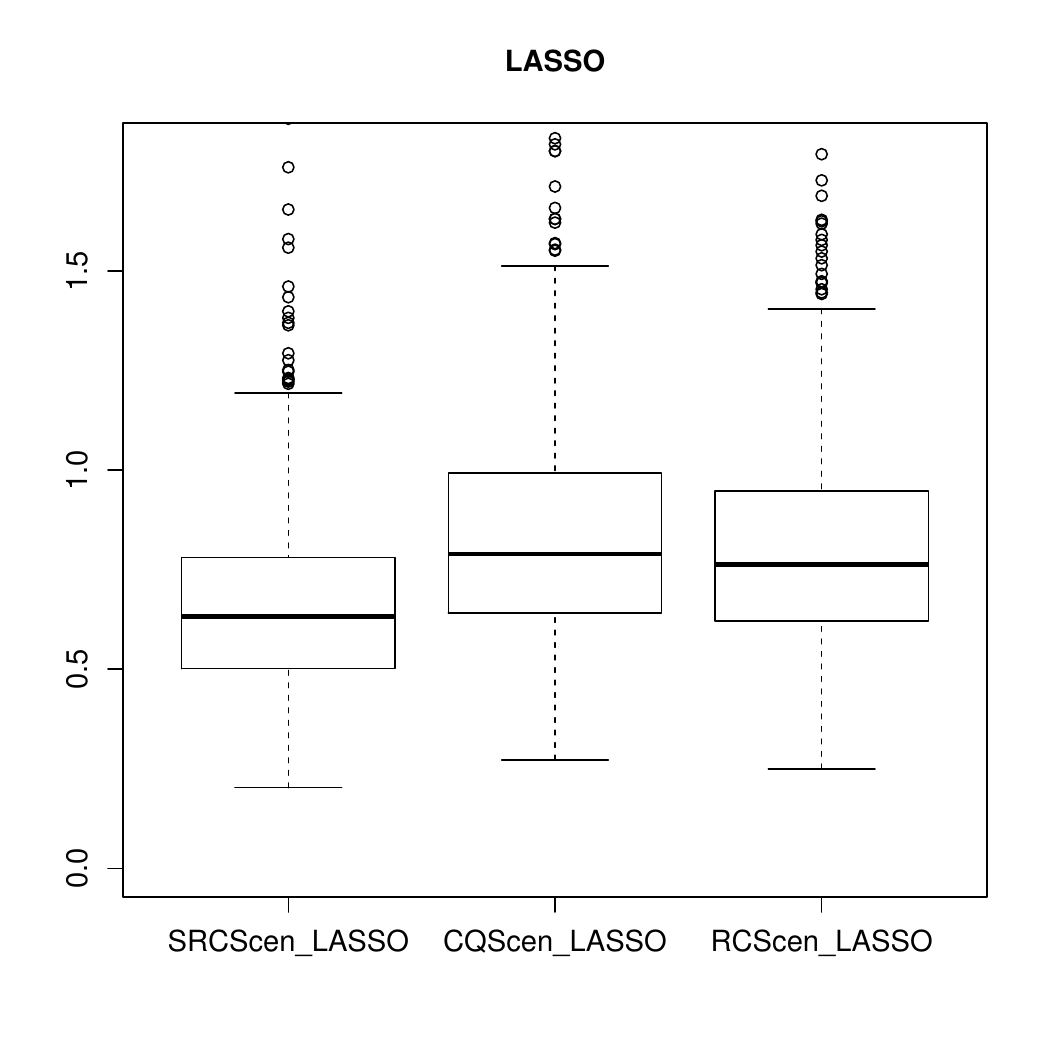}
  \caption{LASSO}
  \label{fig103}
\end{subfigure}
\caption{Boxplots of averaged squared prediction errors.
}
\label{fig10}
\end{figure}

\section{Conclusion}

In this article, to address the new challenges from ultrahigh-dimensional data in robustness and the presence of censoring, especially under a high censoring rate, we  have proposed a new model-free  screening procedure to improve screening efficiency, and established its sure-screening and rank-consistency properties under very weak regularity conditions. The proposed screening approach is invariant under the monotone
transformation and is
 robust under the presence of heavy-tailed distributions,
outliers, and dependent covariates. In particular, the new screening method still works well when a response variable is observed under a high censoring rate.
The superior performance of the proposed screening approach over the existing  methods has been demonstrated through simulation studies and real data analysis.

\section*{References}
\nobibliography*
\noindent\bibentry{chang2013marginal}.\\[.2em]
\bibentry{chen2018reweighted}.\\[.2em]
\bibentry{chen2008extended}.\\[.2em]
\bibentry{chen2018robust}.\\[.2em]
\bibentry{chen2019note}.\\[.2em]
\bibentry{cui2015model}.\\[.2em]
\bibentry{fan2011nonparametric}.\\[.2em]
\bibentry{fan2008sure}.\\[.2em]
\bibentry{fan2009ultrahigh}.\\[.2em]
\bibentry{fan2010sure}.\\[.2em]
\bibentry{he2013quantile}.\\[.2em]
\bibentry{huang2008asymptotic}.\\[.2em]
\bibentry{huang2010variable}.\\[.2em]
\bibentry{leng2007path}.\\[.2em]
\bibentry{li2020nonparametric}.\\[.2em]
\bibentry{li2012robust}.\\[.2em]
\bibentry{li2012feature}.\\[.2em]
\bibentry{lin2013high}.\\[.2em]
\bibentry{liu2018quantile}.\\[.2em]
\bibentry{lin2018model}.\\[.2em]
\bibentry{liu2018new}.\\[.2em]
\bibentry{lo1986product}.\\[.2em]
\bibentry{mai2015fused}.\\[.2em]
\bibentry{martinussen2009covariate}.\\[.2em]
\bibentry{miller1982regression}.\\[.2em]
\bibentry{song2014censored}.\\[.2em]
\bibentry{tang2018exponentially}.\\[.2em]
\bibentry{tang2019feature}.\\[.2em]
\bibentry{tibshirani1997lasso}.\\[.2em]
\bibentry{wang2009forward}.\\[.2em]
\bibentry{wang2008doubly}.\\[.2em]
\bibentry{wu2015conditional}.\\[.2em]
\bibentry{xie2020category}.\\[.2em]
\bibentry{yan2018fused}.\\[.2em]
\bibentry{zhang2007adaptive}.\\[.2em]
\bibentry{zhang2017correlation}.\\[.2em]
\bibentry{zhang2018censored}.\\[.2em]
\bibentry{zhao2012principled}.\\[.2em]
\bibentry{zhu2011model}.\\[.2em]
\bibentry{zhou2021category}.\\[.2em]
\bibentry{zhou2017model}.\\[.2em]
\bibentry{zou2008note}.\\[.2em]

\nobibliography{aaai23}

\end{document}